\documentclass[aps,prb,twocolumn]{revtex4-1}
\usepackage{graphicx}
\usepackage{bm}
\usepackage{amsmath,amsthm,amssymb}

\usepackage{color}
\begin{document}

\title{Spin interferometry in anisotropic spin-orbit fields}

\author{Henri Saarikoski,$^1$ Andr\'es Reynoso,$^{2,3}$ Jos\'e Pablo Baltan\'as,$^4$ Diego Frustaglia,$^4$ and Junsaku Nitta$^5$}

\affiliation{$^1$RIKEN Center for Emergent Matter Science (CEMS), 2-1 Hirosawa, Wako, Saitama 351-0198, Japan}
\email{e-mail: henri.saarikoski@riken.jp}
\affiliation{$^2$Instituto Balseiro and Centro At\'omico Bariloche, Comisi\'on Nacional de Energ\'ia At\'omica, 8400 Bariloche, Argentina}
\affiliation{$^3$Consejo Nacional de Investigaciones Cient\'ificas y T\'ecnicas (CONICET), Argentina}
\email{e-mail: andres.a.reynoso@gmail.com}
\affiliation{$^4$Departamento de Fisica Aplicada II, Universidad de Sevilla, E-41012 Sevilla, Spain}
\affiliation{$^5$Department of Materials Science, Tohoku University, Sendai 980-8579, Japan}
\date{\today}

%
%
%

\begin{abstract}

Electron spins in a two-dimensional electron gas (2DEG) can be manipulated by spin-orbit (SO) fields originating from either Rashba or Dresselhaus interactions with independent isotropic characteristics.
Together, though, they produce anisotropic SO fields with consequences on quantum transport through spin interference. Here we study the transport properties of modelled
mesoscopic rings subject to Rashba and Dresselhaus [001] SO couplings in the presence of an additional in-plane Zeeman field acting as a probe. By means of 1D and 2D quantum
transport simulations we show that this setting presents anisotropies in the quantum resistance as a function of the Zeeman field direction. Moreover,
the anisotropic resistance can be tuned by the Rashba strength up to the point to invert its response to the Zeeman field. We also find that a topological transition
in the field texture that is associated with a geometric phase switching is imprinted in the anisotropy pattern.
We conclude that resistance anisotropy measurements can reveal signatures of SO textures and geometric phases in spin carriers.

\end{abstract}


\maketitle

\section{Introduction}

The electron spin couples directly to magnetic fields via the Zeeman
effect and indirectly to electric fields via the spin-orbit (SO)
interaction, that is due to a special relativistic effect: an electric
field in the laboratory frame gets a magnetic component in the moving
electron's rest frame due to relativistic corrections, which couples
to the electron spin. In mesoscopic quantum wells made of III–-V
zinc-blende semiconductors two different types of SO interaction
arise. The asymmetric potential of the quantum well gives rise to the
Bychkov-Rashba SO interaction via structure inversion
asymmetry.\cite{bychkov} In addition, bulk inversion asymmetry of the
crystal structure leads to the Dresselhaus SO
interaction.\cite{dresselhaus} The magnitude of the SO fields may be
strong, so that effective fields of several teslas can be generated in
semiconductors, which provides effective ways to manipulate spin.
Moreover, the Rashba SO interaction can be tuned with a gate
electrode.\cite{nitta} These properties have made SO fields especially
relevant in studies of spin phenomena.

For instance, the combination of Rashba and Dresselhaus SO
interactions can be used to design complex spin
textures.\cite{schliemann} In this way, when the linear Rashba and
Dresselhaus terms equal each other, an helical spin density wave
(persistent spin helix state) emerges, allowing a huge increase in
spin lifetime \cite{koralek} due to the spin protection by SU(2)
symmetry.\cite{bernevig} Moreover, the possibility of an electric
tuning of the Rashba interaction opens a door to an efficient control
of such spin helix state,\cite{kohda} with potential applications for
nonballistic spin transistors.\cite{schliemann2} However, realistic
implementations are hindered by the complexity of SO fields in
semiconductor heterostructures. 
Besides, applications to several spin phenomena
and SO-induced spin structures require an accurate prediction and
in-situ determination of the SO interaction parameters.\cite{ganichev,sasaki}.

A different prominent feature resulting from the combination of Rashba
and Dresselhaus SO fields in a 2D electron gas (2DEG) is the anisotropic character of electron
transport.\cite{schliemann3}
{ The spin-orbit fields in a 2DEG are associated with weak localization/antilocalization effects.\cite{miller}}
For example, it is known that an in-plane
magnetic field induces a characteristic resistance anisotropy caused
by weak localization.\cite{malshukov} Complex spin-orbit fields in
combination with an in-plane magnetic field are also expected to cause
resistance anisotropy in mesoscopic ring systems due to the influence
of circular interference paths.

In this work, we study electron spin interferometry in mesoscopic rings subject
to $k$-linear Rashba and Dresselhaus SO interactions in the presence
of an external magnetic field applied within the plane of electron
transport. The most notable feature emerging in our calculations is
anisotropy in the resistance as a function of the in-plane magnetic field direction. 
{ In contrast to the 2DEG this anisotropy is associated with the
spin interference around the ring under the influence of spin-orbit fields.\cite{aharonovcasher}
As shown below, time-reversed symmetric Altshuler-Aronov-Spivak (AAS)\cite{altshuler} paths play
a crucial role in the effect in moderate disorder densities.} We predict oscillations in the resistance anisotropy
as a function of both the in-plane and Rashba fields
strengths. Furthermore, we demonstrate signatures of an effective
geometric phase switching in transport anisotropy.

\section{The mesoscopic ring system}

\label{sec:system}

\begin{figure}
\includegraphics[width=\columnwidth]{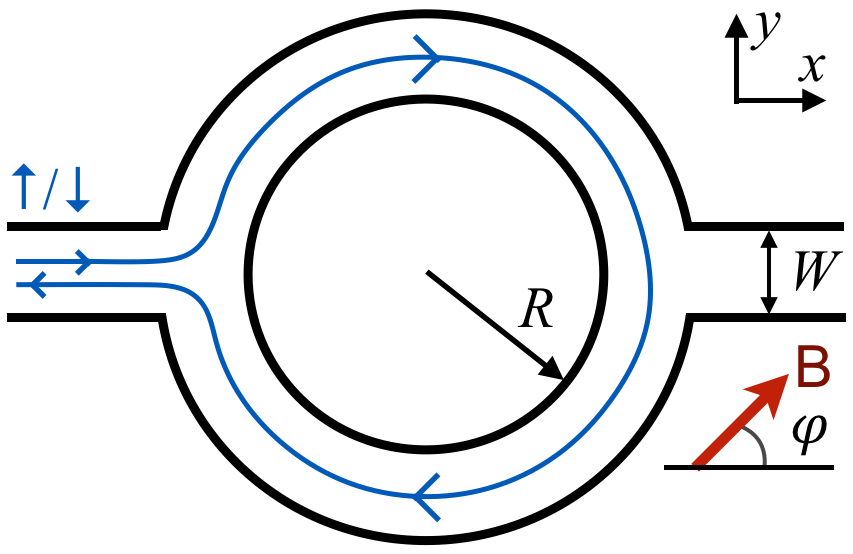}
\caption{Geometry of the 2D ring system. The blue line shows the
  clockwise propagating Altshuler-Aronov-Spivak interference path for
  spins $\uparrow,\downarrow$ that we use in the 1D models. The
  time-reversed counterpart is propagating in the opposite direction.
\label{fig1}}
\end{figure}

Our system consists of a narrow straight mesoscopic wire of width $W$
connected symmetrically to a ring of inner radius $R$ and the same
width, containing a 2DEG. The geometry of the system is depicted in
Fig. \ref{fig1}. The lead orientation angle with respect to the
$x$-axis is denoted by $\omega$. We keep $\omega=0$ in the
calculations, as in Fig.~\ref{fig1}, except when otherwise stated.

Structure and bulk inversion asymmetry results in
the appearance of SO coupling terms. In particular, the linear
Bychkov-Rashba SO interaction may be written as\cite{bychkov}
\begin{equation}
H_{\rm R} = \alpha (k_y {\sigma}_x - k_x {\sigma}_y), 
\label{eq:rashba}
\end{equation}
where ${\bf k}=k_x\hat{\bf x}+k_y\hat{\bf y}+k_{z}\hat{\bf z}$ is the
electron $k$-vector and $\sigma_{x,y,z}$ denote Pauli matrices. In
turn, the Dresselhaus SO interaction depends on the crystal axis
orientation in the quantum well. We consider here the high-symmetry
[001] orientation and neglect $k$-cubic terms. This leads
to\cite{dyakonov,iordanskii}
\begin{equation}
H_{\rm D,[001]} = \beta (k_y {\sigma}_y - k_x {\sigma}_x ).
\label{eq:dresselhaus}
\end{equation}
Here, we are assuming that $\alpha,\beta>0$. Besides, we consider an
in-plane magnetic field
\begin{equation}
{\bf B}(\varphi)=B(\cos\varphi\, \hat{\bf x}+\sin\varphi\, \hat{\bf y})
\label{eq:magfield}
\end{equation}
with $\varphi$ an arbitrary angle characterizing the field direction
with respect to the wire axis (see Fig.~\ref{fig1}). This results in
the Zeeman term
\begin{equation}
H_{\rm Zeeman} = g^* \mu_{\rm B} {\bf B}\cdot {\bm \sigma},
\label{eq:zeeman}
\end{equation}
where ${\bm \sigma}=\sigma_{x}\hat{\bf x}+\sigma_{y}\hat{\bf y}+\sigma_{z}\hat{\bf z}$ is the vector of Pauli matrices, $\mu_{\rm B}$ is the Bohr magneton
and $g^*$ is the effective $g$-factor.

After including the kinetic term, the full Hamiltonian reads
\begin{equation}
H = \hbar^2 k^2/2m^*+H_{\rm R}+H_{\rm D,[001]}+H_{\rm Zeeman},
\label{Hamiltonian}
\end{equation}
where $m^*$ is the effective mass.

\begin{figure}
\includegraphics[width=\columnwidth]{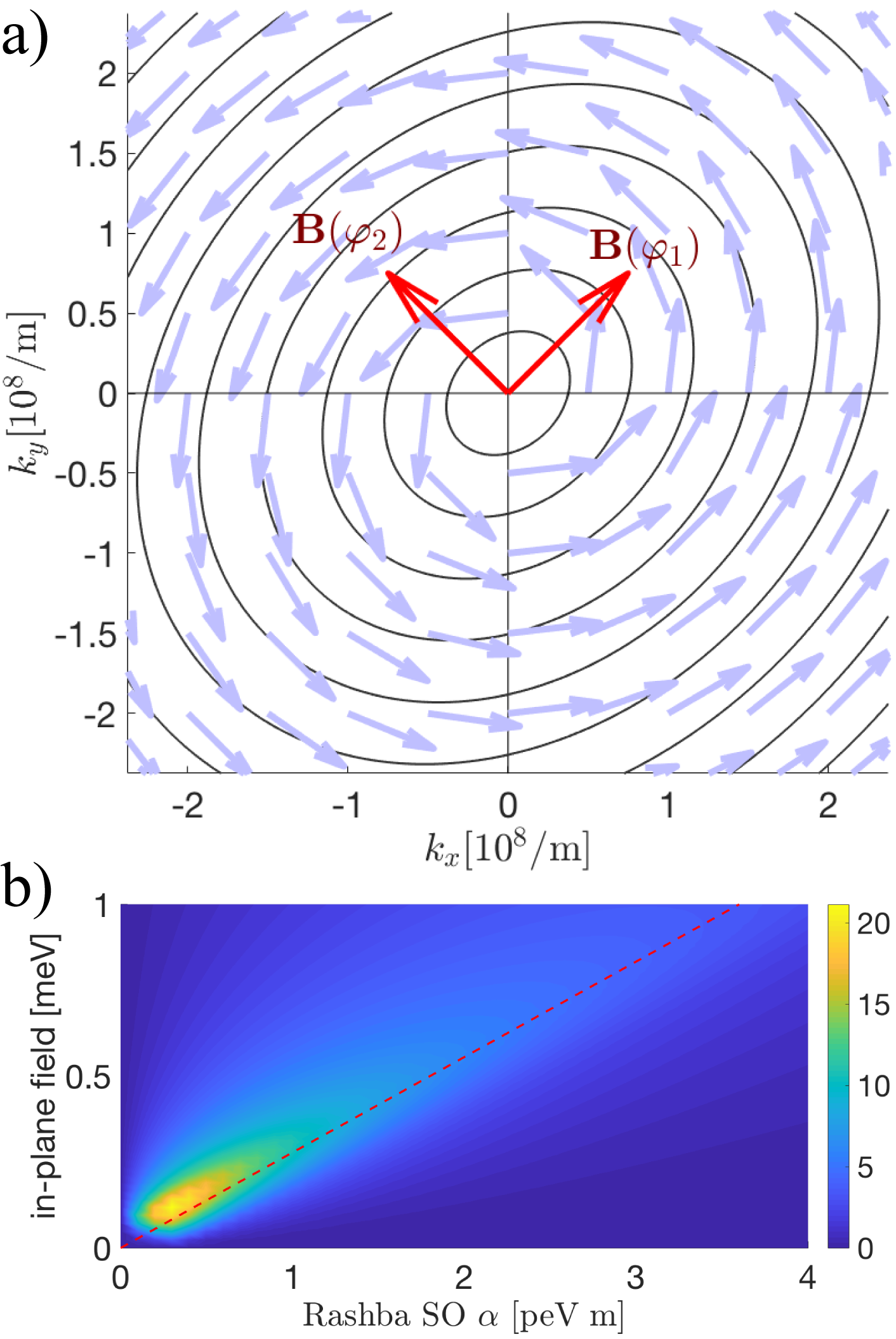}
\caption{a) The combination of Rashba and Dresselhaus spin-orbit
  interactions leads to an anisotropic spin-orbit energy in $k$-space.
  Spin-orbit parameters are here $\alpha=2.64\;{\rm peV\,m}$ and
  $\beta=0.3\;{\rm peV\,m}$.  The figure shows energy countours of the
  spin Hamiltonian (solid lines) and directions of the spin
  eigenstates (arrows).  Angles $\varphi_1=\pi/4$ and
  $\varphi_2=3\pi/4$ correspond to the extreme values of the spin
  orbit field in the ring. b) Anisotropy (in \%) of the total average
  effective field $B_{\rm ave}(\varphi)$ [Eq.~(\ref{avefield})] acting
  on a spin along a circular path around the ring in the 1D
  spin-rotation model. Anisotropy is calculated from $(B_{\rm
    ave}(\varphi=\pi/4)-B_{\rm ave}(\varphi=3\pi/4))/[B_{\rm
      ave}(\varphi=\pi/4)+B_{\rm ave}(\varphi=3\pi/4)]/2$, as a function of the
  Rashba and in-plane field strengths. The Dresselhaus interaction
  energy is fixed to $\beta=0.3\;{\rm peV\;m}$ and a ring radius
  $R=610\,{\rm nm}$ is used.  The dashed line shows where the Rashba
  SO field strength is equal to the in-plane field strength.
\label{figx}}
\end{figure}

The combination of Rashba and Dresselhaus SO interactions leads to 
an anisotropic spin energy in $k$-space\cite{winkler}
(Fig.~\ref{figx}a) experienced by the spin carriers as an anisotropic internal 
magnetic field with extreme values $|\alpha-\beta|k$ and $|\alpha + \beta|k$ along 
the directions $(1,1)/\sqrt{2}$ and $(1,-1)/\sqrt{2}$ in $k$-space, respectively. The anisotropic spin energy can be probed 
by introducing an external Zeeman field coplanar with the ring, giving rise to
anisotropic transport effects with respect to the Zeeman field direction
$\varphi$.

\section{1D and 2D methods}

An appropriate treatment of multiple-mode transport in the wire
requires the use of full 2D simulations. Here, we use the Kwant
code\cite{groth} to implement a tight-binding type Hamiltonian in a 2D
grid and calculate the conductance through the system. We use values
of the material parameters compatible with those of InGaAs, with $m^*
= 0.05m_0$, where $m_0$ is the bare electron mass, and $g^*=3$. An
in-plane field strength of $B=1\;{\rm T}$ corresponds then to 0.17
meV. We keep the Fermi energy at 47.7 meV in all the calculations,
which gives an electron density of approximately $n=1.0\times
10^{12}/{\rm cm}^2$ in the 2DEG at Fermi wavenumber $k_{\rm  F}={2.46}\times 10^8/{\rm m}$. 
The ring inner radius $R$ is fixed to
610 nm, while the wire width $W$ is adjusted between 42 nm and 68 nm,
the former giving 3 conducting modes and the latter giving 5 of them.

Disorder is introduced in the system by adding a spin-independent
random scattering potential in the lattice.\cite{ando} { The resulting conductance is strongly dependent on the particular
  realization of this random potential, hence we perform disorder
  averaging in our simulations, giving the statistical accuracy of our
  results in terms of $\pm\sigma$ errorbars. In experiments, this may
  be done effectively by constructing a self-averaging sample
  consisting in a network containing a large number of
  rings.\cite{nagasawa2}. In this setup, AAS paths are especially
  important\cite{richter} since direct interference paths are
  strongly affected by small sample asymmetries, disorder, scattering
  between the transport modes as well as energy averaging. For AAS
  time-reversed symmetric paths, on the contrary, the acquired phase $k_F2\pi R$
  between clockwise and anticlockwise winding paths is cancelled out
  (in contrast to the spin phase difference), so the effects of disorder
  are minimized. With this in mind, we use a disorder mean-free path of $\ell=2.5\;{\mu}{\rm m}$ unless otherwise stated.
This is shorter than the inner circumference of the ring, $3.8\; {\mu}{\rm  m}$, and comparable to
the estimated mean free path in the array of semiconductor rings in Ref. \onlinecite{nagasawa2}.
This moderate disorder density favours the AAS interference paths  (see Fig.~\ref{fig1}) prevailing in the experiments.}

In addition, we apply approximate 1D models to gain physical insights
into the spin dynamics and, in particular, into the origin of the
anisotropy in resistance reported below. 
We neglect curvature and torsion effects that arise due to the steep curvature at the
intersection.\cite{ortix,ying} On the one hand, we use a 1D
tight-binding-based approach that incorporates the full Hamiltonian
in a chain (see Appendix \ref{appendixtb} for details). On the other
hand, we introduce a 1D spin-rotation model that neglects the orbital
part of the Hamiltonian. { In this model we assume moderate disorder density in the system
and focus therefore on the AAS interference paths}.
Thus, the total probability of backscattering is
calculated from the reflected wave at the intersection and waves
transmitted around the ring from
\begin{equation}
\psi = p_1 \psi_0+ p_2(\psi_++\psi_-),
\label{backscattering}
\end{equation}
where the spinor states $\psi_{+,-}$ correspond to clockwise ($+$) and
anticlockwise ($-$) moving spins, $\psi_0$ stands for directly
reflected spins, and $p_{1,2}$ are the backscattering amplitudes for
directly reflected spins and spins propagating once around the ring,
respectively. As shown in Appendix \ref{appendixspinrot}, the
resistance can be expressed in a closed form in terms of $\psi$.

In the following, we use two independent characterizations of the
anisotropy observed in the resistance $R_{\Omega}(\varphi)$ across the ring. Resistance
anisotropy between fixed axes corresponding to different in-plane
orientations $\varphi_A$ and $\varphi_B$ is calculated from
\begin{equation}
A_{\rm R}(\varphi_A,\varphi_B)=\frac{ R_{\Omega}(\varphi_A)-R_{\Omega}(\varphi_B)}{[R_{\Omega}(\varphi_A)+R_{\Omega}(\varphi_B)]/2}
\label{angleanisotropy}
\end{equation}
by using the methods described above. In addition, when possible, we compute the angle-averaged
resistance $\langle R_{\Omega}(\varphi) \rangle=\int_0^\pi R_{\Omega}(\varphi)\,{\rm d}\varphi/\pi$, from which we calculate the $\varphi$-dependent resistance anisotropy
\begin{equation}
A_{\rm R}(\varphi)=\frac{R_{\Omega}(\varphi)-\langle R_{\Omega}(\varphi) \rangle}{\langle R_{\Omega}(\varphi) \rangle}.
\label{averagedanisotropy}
\end{equation}
Resistance is given here in units of $1/G_0$, where $G_0 =e^2/h$ is the
quantum of conductance.

\section{Transport anisotropy}

\begin{figure}
\includegraphics[width=\columnwidth]{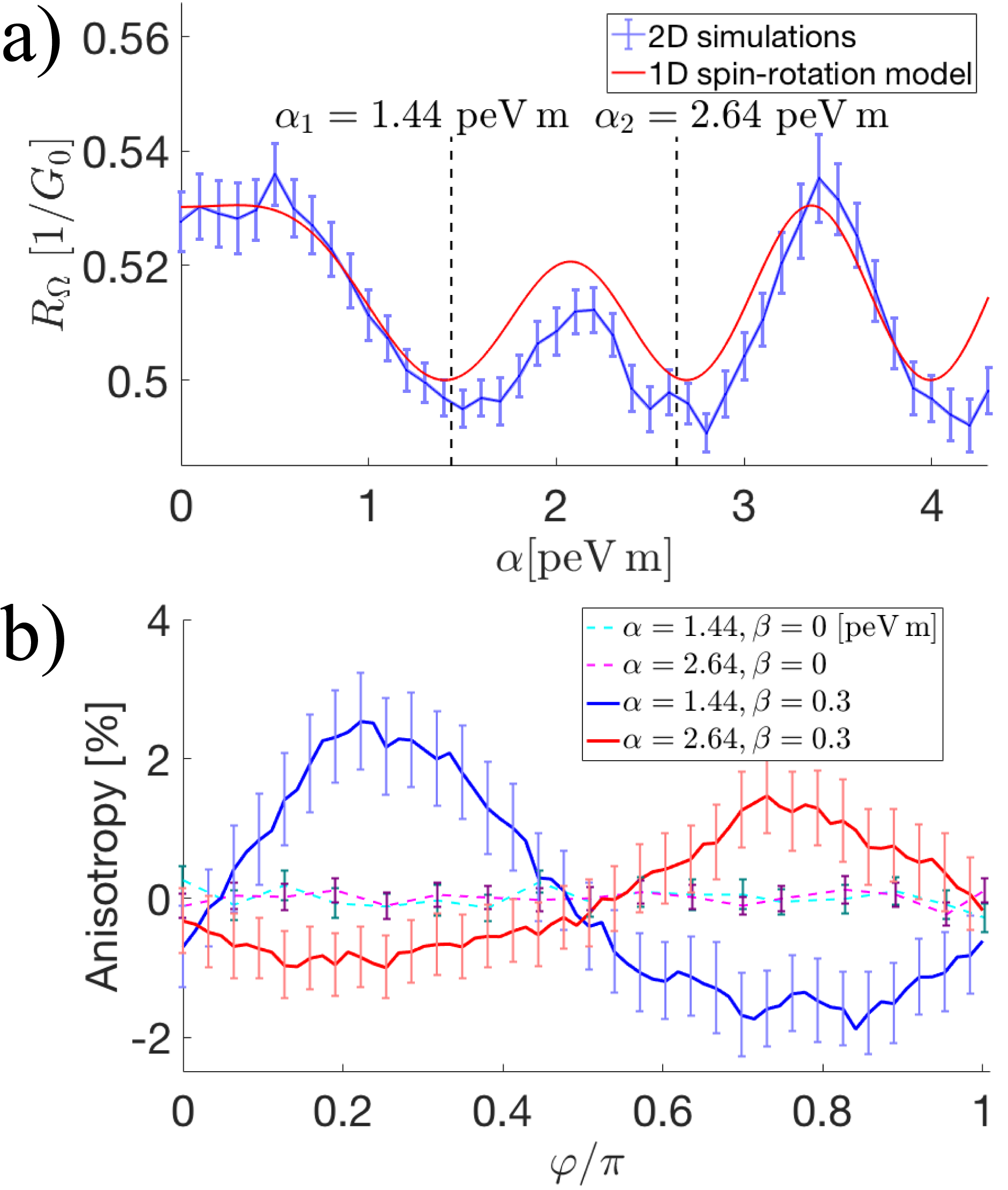}
\caption{a) Resistance in a 3-mode ring system as a function of Rashba
  SO coupling $\alpha$ and at fixed Dresselhaus SO coupling
  $\beta=0.3\,{\rm peV\,m}$ in the absence of an external magnetic
  field (${\bf B}=0$).  The results are calculated using the 2D model
  and compared to those obtained with the single-mode 1D spin-rotation
  model with AAS paths ($p_1=0$, $p_2=0.12$). The Rashba couplings
  $\alpha_{1,2}$ shown in the figure coincide approximately with the
  first resistance minima. b) Resistance anisotropy $A_{\rm R}(\varphi)$
  [Eq.~(\ref{averagedanisotropy})] corresponding to an in-plane Zeeman field
  strength of $0.17\,{\rm meV}$ and Rashba SO fields
  $\alpha_1=1.44\,{\rm peV\,m}$ and $\alpha_2=2.64\,{\rm peV\,m}$. The
  Dresselhaus field strength is fixed to $\beta=0.3\,{\rm peV\,m}$.
  For comparison, we include the case $\beta=0$ showing no
  anisotropy (dashed lines), see Appendix~\ref{isotropiccase}.
\label{fig2}}
\end{figure}

In the following we focus on anisotropic SO fields\cite{footnote}
where the Dresselhaus SO interaction is low, {\em i.e.}, $\beta \ll
\alpha$.  We consider first the case of zero in-plane magnetic field,
${\bf B}=0$.  Aharonov-Casher resistance oscillations calculated as a
function of the Rashba spin-orbit coupling $\alpha$ in a disordered
3-mode ring system are shown in Fig.~\ref{fig2}a for a fixed
Dresselhaus SO coupling $\beta=0.3\,{\rm peV\,m}$, close to the
Dresselhaus interaction strength found in InGaAs
heterostructures.\cite{sasaki} { We point out that the
  oscillation frequency of the resistance as a function of $\alpha$ is
  twice the frequency of the oscillations due to direct interference
  paths, which become dominant in the clean limit of ballistic
  transport. This allows us to associate the observed behavior with
  AAS interference paths. In fact, the results are} adequately
reproduced by the 1D single-mode spin-rotation model with AAS paths
for an appropriate choice of the scatterig amplitudes ($p_1=0$, $p_2=0.12$).

We add now an external in-plane magnetic field to the SO field. The
average effective field strength acting on a spin in a round-trip
around the ring becomes then anisotropic. This can be illustrated by
computing the magnitude of the average effective field for a one-dimensional
single-channel ring as a function of $\varphi$ (see Appendix~\ref{appendixspinrot})
\begin{equation}
B_{\rm ave}(\varphi)=\int_0^{2\pi} |{\bf B}_{\rm R}+{\bf B}_{\rm D}+g^* \mu_{\rm B}{\bf B}_{\rm in-plane}(\varphi)|d\theta/2\pi,
\label{avefield}
\end{equation}
and comparing its values for two different orientations of the
in-plain field. This is depicted in Fig.~\ref{figx}b. Notice that the
field anisotropy is largest when all the field components are of
comparable magnitude.

\begin{figure}
\includegraphics[width=1\columnwidth]{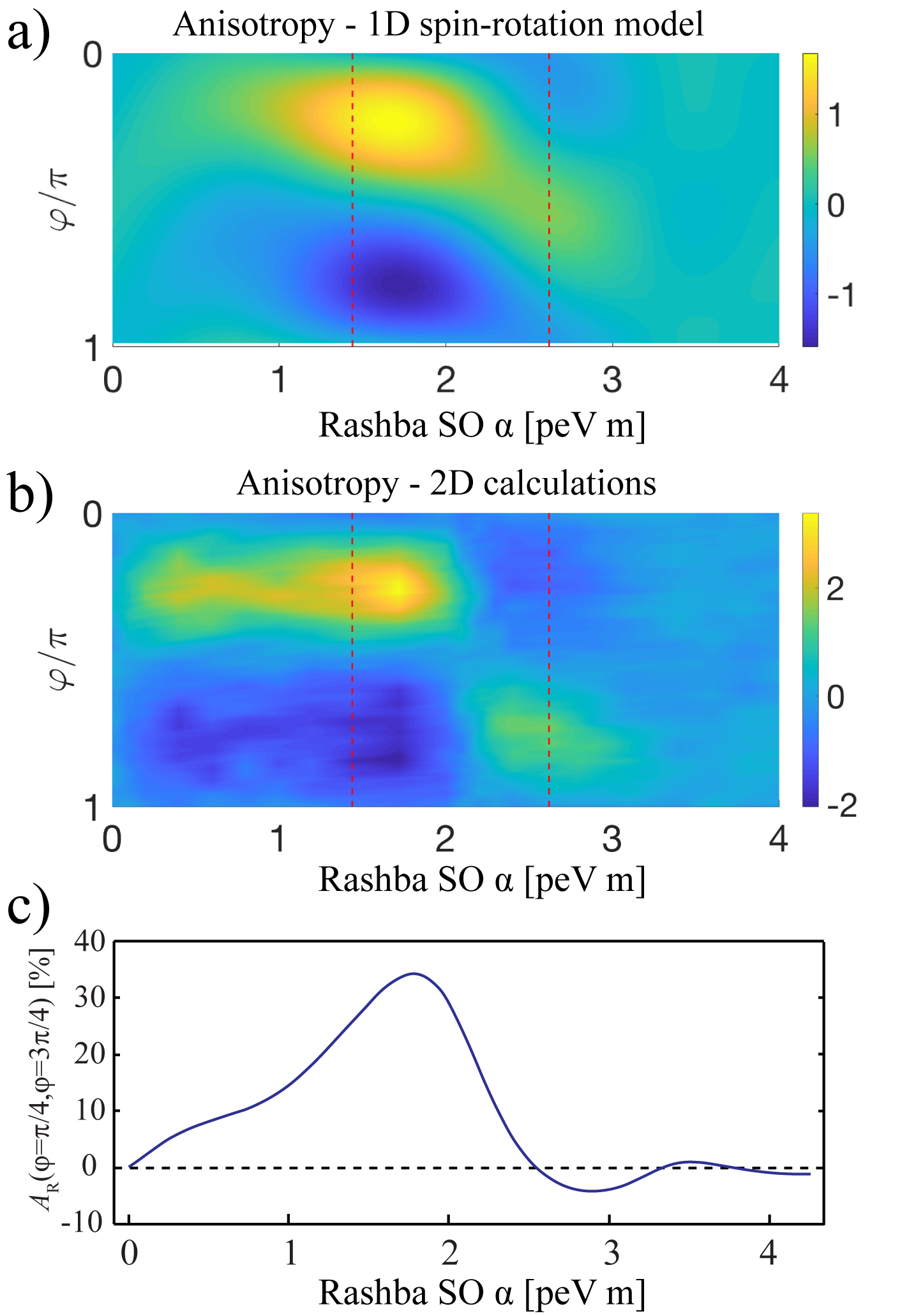}
\caption{Resistance anisotropy $A_{\rm R}(\varphi)$
[Eq.~(\ref{averagedanisotropy})], in \%, for fixed Dresselhaus
($\beta=0.3\,{\rm peV m}$) and in-plane ($0.17\,{\rm meV}$) magnetic
fields as a function of the Rashba spin-orbit coupling $\alpha$ and
the in-plane field direction $\varphi$. Panels (a) and (b)
correspond to the results obtained with the 1D spin-rotation model
and full 2D transport simulations, respectively. The 1D model uses
backscattered AAS paths ($p_1=0$, $p_2=0.12$). In both cases, a
``phase switch'' in anisotropy is observed close to
$\alpha=2.1\,{\rm peV\, m}$. The dashes lines indicate the Rashba SO
couplings $\alpha_{1,2}$ shown in Fig.~\ref{fig2}. 
c) Resistance anisotropy $A_{\rm R}(\varphi=\pi/4,\varphi=3\pi/4)$
calculated with the 1D tight-binding method demonstrates an oscillating
phase in anisotropy. Anisotropy is calculated from the AAS paths
with the Dresselhaus SO strength $\beta=0.3\;{\rm peV\,m}$,
and in-plane field strength 0.17 meV. Isotropic background conductance in
the calculation is fixed to $G_{\rm B}=2.5G_0$ (see Appendix~\ref{appendixtb}). 
\label{fig-anisotropyaxis-1D}}
\end{figure}

The anisotropic nature of the effective field manifests as a
anisotropy in the resistance across the ring. In particular, for fixed
Rashba, Fig.~\ref{fig2}b shows a non-zero resistance anisotropy
$A_{\rm R}(\varphi)$, strongly dependent on the external field
orientation. Likewise, an unambiguous oscillation of the resistance
anisotropy as a function of $\alpha$ is observed when the direction of
the external field remains fixed. This is most visible in
Figs.~\ref{fig-anisotropyaxis-1D}a and b, showing
the resistance anisotropy as a function of the field orientation
$\varphi$ and the Rashba SO strength $\alpha$ for fixed
$\beta=0.3\,{\rm peV\; m}$, calculated using 2D simulations and the 1D
spin-rotation model. { Degree of anisotropy decreases with increasing $\alpha$ since the field anisotropy
also decreases (Fig.~\ref{figx}b).}

Figures \ref{fig-anisotropyaxis-1D}a and b show a change in the
sign of phase of resistance anisotropy at $\alpha=2.1\,{\rm peV\, m}$
and again at about $3.3\,{\rm peV\, m}$. Using definition (\ref{angleanisotropy})
this is described by the oscillating sign of $A_{\rm R}(\varphi=\pi/4,\varphi=3\pi/4)$
as a function of $\alpha$. Figure~\ref{fig-anisotropyaxis-1D}c shows oscillations
in $A_{\rm R}(\varphi=\pi/4,\varphi=3\pi/4)$ as reproduced by the 1D tight-binding approach.
Anistropy changes sign here at $\alpha=2.5\,{\rm peV\, m}$, $3.3\,{\rm peV\, m}$ and $3.8\;{\rm  peV\, m}$.
We note that the overall behavior in the 1D tight-binding approach is consistent with 
the 1D spin rotation model as well as the 2D calculations in Figs.~\ref{fig-anisotropyaxis-1D}a and b.

{
In addition to resistance anisotropy originating from AAS spin interference 
multi-mode rings are expected to display resistance anisotropy caused by weak localization/antilocalization in the multi-mode wire.\cite{malshukov}.
However, the resistance anisotropy due to the latter contribution reverses sign only when either the Rashba
or Dresselhaus coupling parameter changes its sign.  This is in contrast to our results that show oscillations in the sign of
$A_{\rm R}(\varphi=\pi/4,\varphi=3\pi/4)$ with increasing spin-orbit field (Fig.~\ref{fig-anisotropyaxis-1D}).
We find that this pattern of anisotropy sign reversal persists at moderate disorder densities with $0.45\;{\rm \mu  m}<\ell<3.0\;{\rm \mu m}$
(see Appendix~\ref{supplydata}).

We find no sign reversal in $A_{\rm R}(\varphi=\pi/4,\varphi=3\pi/4)$ in the very disordered regime $\ell < 0.15\;{\rm \mu  m}$.
Resistance anisotropy originates then mainly from the weak localization/antilocalization effects
in the multi-mode wire (see Appendix~\ref{supplydata}).
On the other hand, in the regime of weak disorder $\ell>3.0\;{\rm \mu m}$ spin interference from
direct path Aharonov-Casher contributions start to dominate changing the pattern of resistance anisotropy.
These calculations indicate that AAS ring interference is the most important source of resistance anisotropy in the 2D simulations
at moderate disorder densities. 
}


\begin{figure}
\includegraphics[width=1\columnwidth]{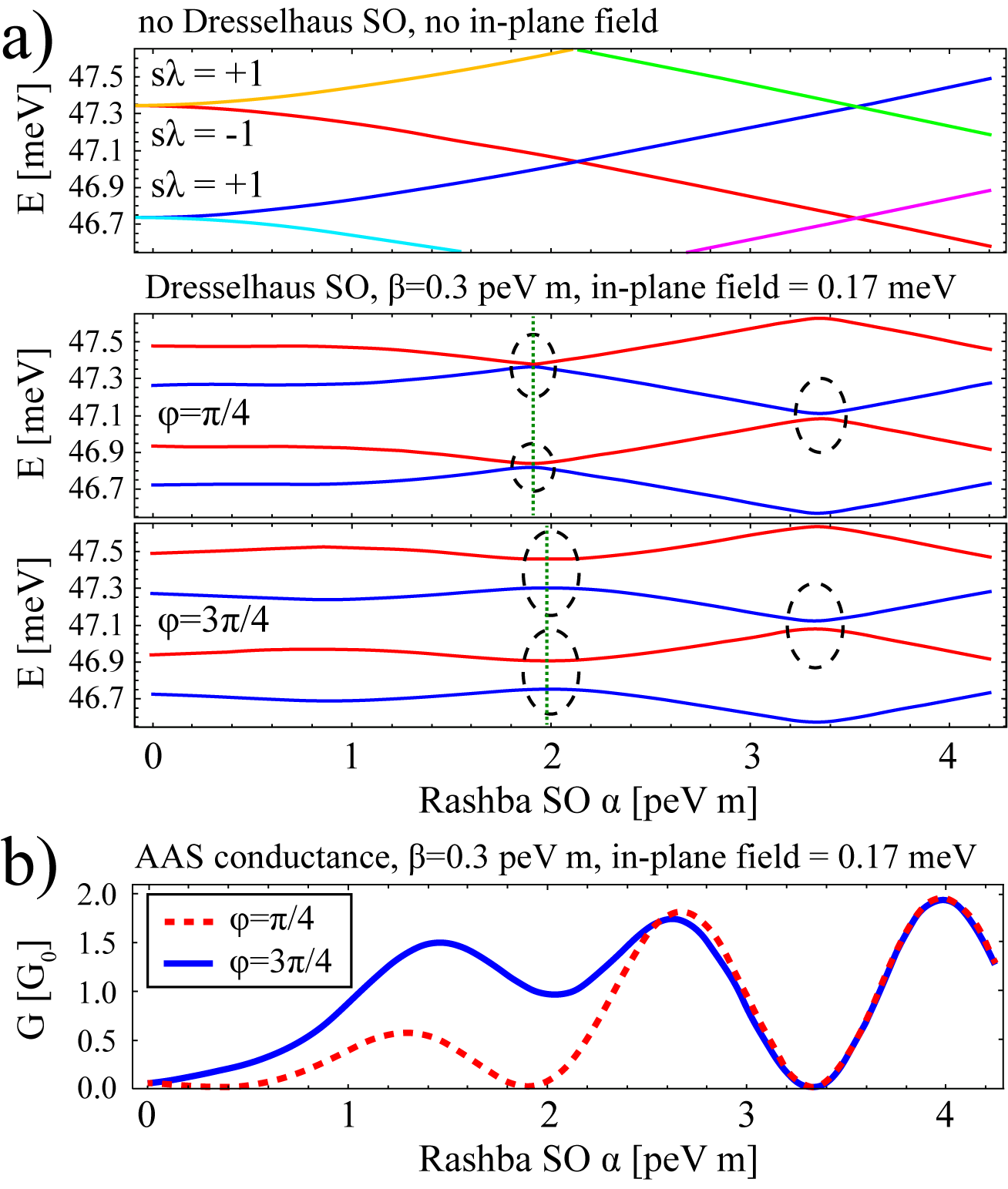}
\caption{a) Eigenstate
  spectrum of an isolated spin-orbit ring system as a function of the
  Rashba spin-orbit coupling $\alpha$ in the cases of vanishing
  Dresselhaus and in-plane fields (upper panel) and finite Dresselhaus
  and in-plane fields, $\beta=0.3\;{\rm peV\;m}$ and $B=0.17\;{\rm
    meV}$, for two in-plane orientations, $\varphi=3\pi/4$ (middle
  panel) and $\varphi=\pi/4$ (lower panel).  Anti-crossing gaps appear
  in the spectrum. Gap sizes (circles) and positions (dashed green lines) depend on the in-plane
  field direction. The results are computed using the
  1D tight-binding model. b) Corresponding AAS anisotropic conductance at $\varphi=\pi/4$ and
  $\varphi=3\pi/4$.
\label{tb-supply}}
\end{figure}


\subsection{Anisotropy in the ring spectrum}

To clarify the origin of the resistance anisotropy we study the
eigenstates of an isolated single-mode ring. In the absence of both
Dresselhaus and in-plane fields, an exact expression for the energy
levels corresponding to the different eigenstates can be found
\cite{frustaglia}, which reads
\begin{equation}
E=\frac{\hbar^2}{2m^* R^2}\Big[\Big(\lambda n
+\frac{1}{2}\Big)^2+\frac{1}{4}+s\Big|\lambda n +\frac{1}{2}\Big|\sqrt{1+\Big(\frac{2\alpha m^* R}{\hbar^2}\Big)^2}\Big],
\end{equation}  
where $\lambda=\pm 1$ denotes the rotation direction of the electron
in the ring, $s=\pm 1$ its spin, and $n > 0$ is the (integer) orbital
quantum number. For a vanishing SO field ($\alpha=0$), the eigenstates
show a 4-fold degeneracy: a 2-fold Kramers degeneracy associated with the rotation
direction around the ring and a 2-fold spin degeneracy. Spin
degeneracy is lifted by the Rashba spin-orbit interaction (upper panel
in Fig.~\ref{tb-supply}a). Eventually, as the Rashba
strength increases new level crossings show up and the 4-fold
degeneracy is restored, but this time involving states of different
$n$, $\lambda$, and $s$, which carry a different total phase. An
example is shown in the upper panel of
Fig.~\ref{tb-supply}a where a 4-fold degeneracy appears
near $\alpha=2.2\;{\rm peV\;m}$ and $E=47.1\;{\rm meV}$ for states
corresponding to $(n=143,s=+1,\lambda=+1)$, $(n=144,s=-1,\lambda=-1)$,
$(n=145,s=-1,\lambda=+1)$, and $(n=146,s=+1,\lambda=-1)$ with a
round-trip phase difference of $2\pi$ and $6\pi$ for $s=-1$ and $s=+1$
spin species, respectively. This is associated with the resistance
maximum for AAS transport due to constructive backscattering
interference of the same spin species.

The presence of an anisotropic SO field ($\beta\ne 0$) and an in-plane
Zeeman field modifies the previous description substantially. In
particular, anti-crossing gaps open (dashed lines in the lower panel
of Fig.~\ref{tb-supply}a). This is detrimental to
the constructive interference of the AAS paths and, consequently, the conductance
at the gap position increases. Most importantly, both the
gap sizes and positions depend on the in-plane field direction, from
which an anisotropy in conductance naturally emerges. { This is
demonstrated in Fig.~\ref{tb-supply}b
for two different orientations of the Zeeman field $\varphi=\pi/4$ and $\varphi=3\pi/4$: The larger gap sizes appearing at slightly different values of $\alpha$ for $\varphi=3\pi/4$ result in a higher AAS conductance 
with shifted minima when compared with the case $\alpha=\pi/4$}. In addition, the sizes of the gaps appearing in
the anisotropic spectrum decreases with $\alpha$ and, correspondingly,
so does the anisotropy.


\subsection{Zeeman oscillations in anisotropy}

\begin{figure}
\includegraphics[width=\columnwidth]{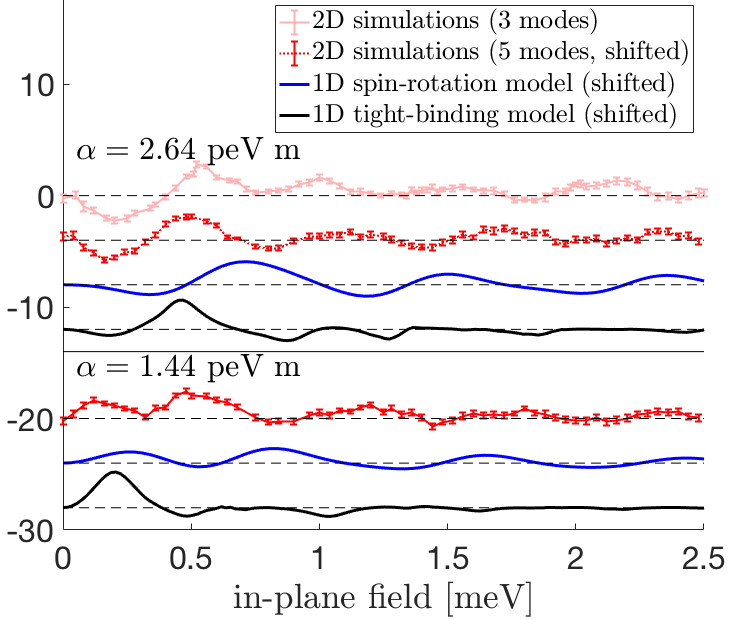}
\caption{Zeeman oscillations in resistance anisotropy of a mesoscopic
  ring calculated with 1D and 2D methods at Rashba SO coupling
  strength $\alpha=2.64\;{\rm peV\, m}$ (above) and $\alpha=1.44\;{\rm
    peV\, m}$ (below).  Anisotropy (in \%) is calculated between
  $\varphi_1=\pi/4$ and $\varphi_2=3\pi/4$ axis from $A_{\rm
    R}(\varphi_1,\varphi_2)$ [Eq.~(\ref{angleanisotropy})].  2D
  simulations were performed for a 3-mode system (at the higher Rashba
  field) as well as for a 5 -mode system (both for lower and higher
  Rashba fields). The isotropic contribution to conductance in the 1D
  tight-binding method is fixed to $G_{\rm B}=5G_0$ (see Appendix \ref{appendixtb}).  The Dresselhaus
  SO coupling is fixed at $\beta=0.3\,{\rm peV\, m}$ in all
  calculations.  The mean free path of electrons in the 2D method is
  $\ell=2.5\;\mu {\rm m}$ and the ring radius is $0.61\;\mu{\rm m}$.
\label{highfieldfig}}
\end{figure}

The oscillations in resistance anisotropy as a function of the Rashba parameter $\alpha$
are correlated with phases acquired by the spins when travelling along the ring, with
contributions of both dynamic and geometric origin. Therefore anisotropy oscillations
are expected also as a function of the in-plane field strength.
Fig.~\ref{highfieldfig} shows Zeeman oscillations in the resistance
anisotropy $A_{\rm R}(\varphi_1=\pi/4,\varphi_2=3\pi/4)$ obtained by
using the 2D multi-mode method, the 1D tight-binding method, and the
1D spin-rotation model.
The oscillations in resistance originate partially from the spin interference between the directly
reflected path and the AAS paths {(see Appendix \ref{appendixzeeman} for further discussion)}. This is modeled by the finite $p_1$ term in the 1D spin-rotation model.
To account for disorder effects in the 1D spin-rotation model we choose an amplitude
$p_1=0.06$ for the directly reflected path and $p_2=0.1$ for the AAS paths.
A weak Dresselhaus component $\beta=0.3\;{\rm  peV\;m}$
is considered in the two cases displayed in the figure,
corresponding to Rashba field intensities $\alpha=2.64\;{\rm peV\;m}$
and $\alpha=1.44\;{\rm peV\;m}$. We note that all the three methods
give oscillation periods of about 0.7 meV.


\section{Imprints of geometric phase switching}

\label{sec:topo}
In the last years, the relevance of geometric spin phases has been
strikingly exemplified in studies on the magnetoconductance of
mesoscopic arrays of rings with strong SO coupling.\cite{nagasawa}
This includes the independent manipulation of geometric spin phases by
means of weak external in-plane magnetic fields.\cite{nagasawa2}
Interestingly, a proposal to use an external in-plane field for
switching the effective field's topology in SO ring interferometers
within the adiabatic regime, causing an observable imprint in the
conductance, goes back to a work by Lyanda-Geller\cite{lyanda} in the
early '90s (see Ref.~\onlinecite{ghahari} for a recent observation of
Berry phase\cite{berry} switching in graphene resonators). Still, the
adiabatic treatment proposed by Lyanda-Geller appears to be
inadequate, which has motivated the search of corresponding effects
for nonadiabatic spin dynamics.\cite{anandan} Recently, this has led
to the identification of a so called effective geometric
phase\cite{saarikoski} governing a topological transition in the
nonadiabatic spin dynamics correlated with a topological transition in
the field texture. It has been suggested that this effective phase of
geometrical origin could be studied by either interferometric means in
ring systems or analyzing resonances in spin or other two-level
quantum systems.\cite{saarikoski2,reynoso,baltanas} Here we show that
signatures of a transition sharing some of the features associated to
the effective geometric phase may be observed in the anisotropy
oscillations considered in this work.

\begin{figure*}
\includegraphics[width=1.6\columnwidth]{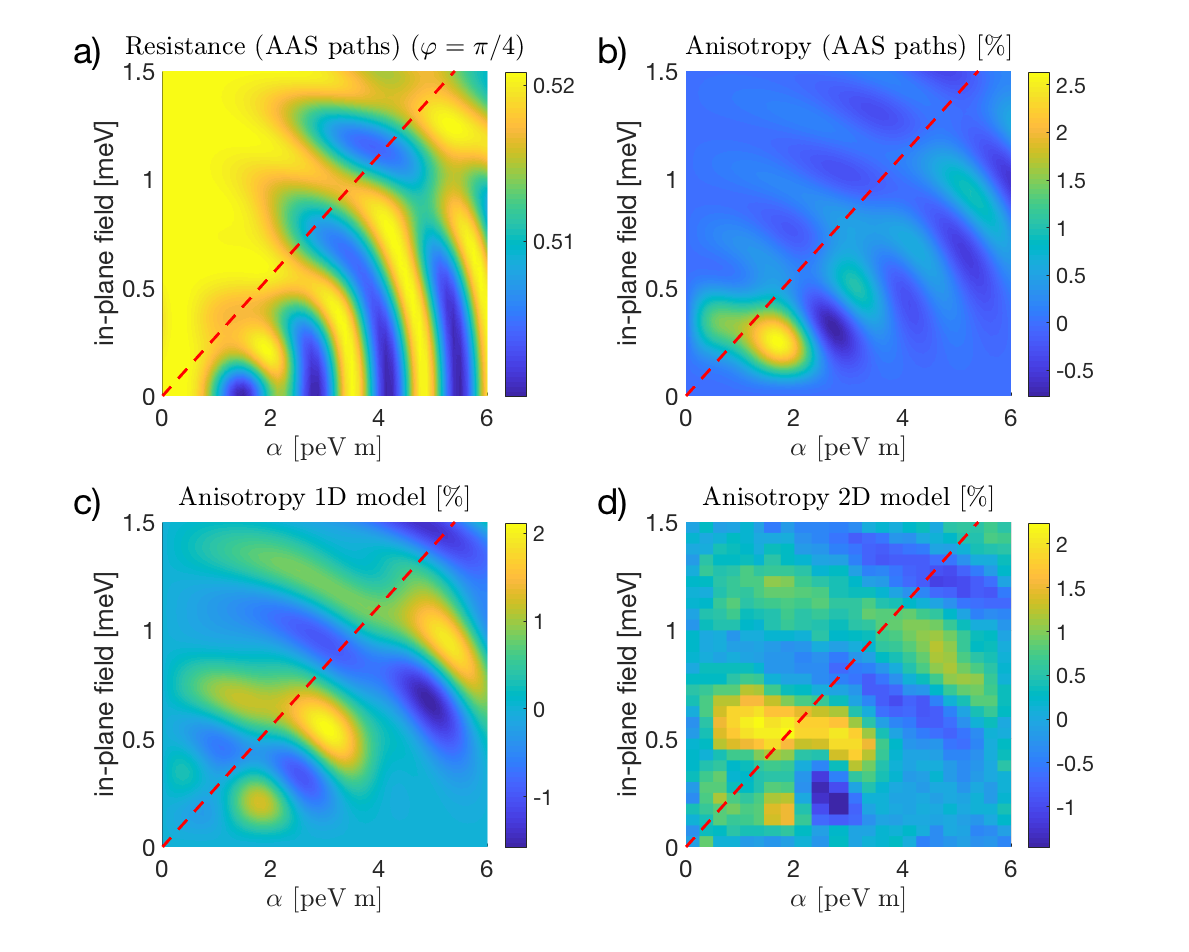}
\caption{a) Resistance calculated from the backscattered AAS paths (1D spin-rotation model with $p_1=0$, $p_2=0.12$) 
for an in-plane field angle $\varphi_1=\pi/4$.
b) Resistance anisotropy for AAS paths averaged over lead orientation $\omega$ in the 1D spin-rotation model.
Note the phase shift in the anisotropy oscillations near the critical line (dashed) where the Rashba SO field strength is equal to the in-plane field strength. 
c) Resistance anisotropy calculated using the 1D spin-rotation model with a direct backscattering amplitude $p_1=0.06$ added.
d) Resistance anisotropy calculated using the 2D model for mean-free path of $\ell=2.5\;{\mu}$m and $\omega=0$.
Dresselhaus SO $\beta=0.3\;{\rm peV\,m}$ in all calculations. All anisotropies are calculated between in-plane field directions $\varphi_1=\pi/4$ and $\varphi_2=3\pi/4$ using $A_{\rm R}(\varphi_1,\varphi_2)$ [Eq.~(\ref{angleanisotropy})].
\label{figy}}
\end{figure*}

1D and 2D numerical results show a characteristic shift in the pattern of
oscillations of the resistance anisotropy $A_{\rm R}(\varphi_1=\pi/4,\varphi_2=3\pi/4)$
close to the critical line where the in-plane field is equal to the Rashba SO field and the field
topology changes\cite{saarikoski} (see panels b, c an d in
Fig.~\ref{figy}). The strength of the Dresselhaus field is weak
enough ($\beta=0.3\;{\rm peV\,m}$) to ensure that the total SO field is
approximately rotationally symmetric.
We note that the resistance calculated from the backscattering probability for AAS
paths by using the 1D spin rotation model ($p_1=0$, $p_2=0.12$) does
not display a clear evidence of a topological transition due to the
suppression of Zeeman oscillations (Fig.~\ref{figy}a). 

1D results of the spin-rotation model for pure AAS paths are shown in Fig.~\ref{figy}b.
In general, the pattern close to the critical line depends on the
lead orientation due to nonadiabatic effects and the slight anisotropy
of the SO field. However, this effect is not important in our case and it
is smoothed out when an average over the lead orientation $\omega$ is performed (data for
$\omega=0$ is shown in Appendix~\ref{supplydata}). Motivated by the
results shown in Appendix~\ref{appendixzeeman}, we analyze the
contribution of directly backscattered electrons in panel c, where
$p_1=0.06$ is used. In this case, the reflected component interferes
with that propagating around the ring and produces an interference
pattern with twice the periodicity observed for AAS paths. The
interference pattern is thus more complex but, nevertheless, displays
the distinct phase dislocation across the critical line.

2D simulations\cite{footnote1} in Fig.~\ref{figy}d show an agreement
with those obtained in the 1D case with the spin-rotation model in
Fig.~\ref{figy}c. It demonstrates that directly reflected spins play a
relevant role in the interference pattern. It is worth to note the
slightly shorter periodicity as a function of the in-plane field
strength in 2D calculations when compared to the 1D case. This is due
to the slightly longer average path around the ring that the electrons
experience in the presence of disorder scattering, resulting in a
higher Zeeman phase.


\section{Conclusions}

We have shown that the interplay between Rashba and Dresselhaus [001] SO fields in mesoscopic rings leads
to transport anisotropies when probed by an inplane Zeeman field. This is due to the anisotropic nature of the total
effective fields that determine the dynamics of (counter)clockwise interfering spin carriers, which also manifests as anisotropic
avoiding crossings in the spectrum of isolated rings tuned by the fields. We found that the resistance anisotropy oscillates as
a function of both Zeeman and SO field strengths. This suggests that field components can be characterized together with dynamic
and geometric spin phases through the measurement of resistance anisotropies. 

Our 2D simulations show that the anisotropic signal is expected to be robust against disorder and other sample asymmetries.
Resistance anisotropy measurements may therefore provide clear evidence of an effective geometric-phase switching in the regime
of nonadiabatic spin dynamics, where a direct observation of such topological transition may be challenging due to spin dephasing
and relaxation in high Zeeman fields.\cite{meijer}

\section*{Acknowledgments}

This work was supported by Japan Society for the Promotion of Science
through Grant-in-Aid for Scientific Research (C) No. 17K05510,
Grant-in-Aid for Specially Promoted Research No. H1505699,
Grant-in-Aid for Scientific Research on Innovative Areas
No. JP15K21717, and by Project No. FIS2014-53385-P (MINECO/FEDER,
Spain). JPB thanks the hospitality of Centro At\'omico Bariloche
(Argentina), where part of this work was done. The 2D simulations were
calculated using the Hokusai system provided by Advanced Center for
Computing and Communication (ACCC) at RIKEN.

\begin{appendix}

\section{Computational methods}

\subsection{1D tight-binding approach}

\label{appendixtb}
The 1D tight-binding method is based on a single channel approach where we apply the customary finite difference method to the Hamiltonian of
Eq. (\ref{Hamiltonian}). The ring is reduced to $N$ sites separated by the lattice spacing $a_0=P/N$, with $P=2\pi R_c$ the ring perimeter, and
site positions $\mathbf{r}_{n}=\left( R_{c}\cos (2\pi (n-1)/N) , R_{c}\sin (2\pi (n-1)/N) \right)$, where $\mathbf{r}_{N+1}=\mathbf{r}_{1}$
in ring geometry. The tight-binding Hamiltonian of the ring becomes
\begin{widetext}
\begin{eqnarray}
  \hat{H}&=&\sum_{n=1}^N \sum_{\sigma\sigma'} \left[\hat{c}_{\mathbf{r}_n\sigma}^\dagger \left(S\right)_{\sigma\sigma'} \hat{c}_{\mathbf{r}_{n}\sigma'}    + \left(\hat{c}_{\mathbf{r}_n\sigma}^\dagger \left(T_n\right)_{\sigma\sigma'} \hat{c}_{\mathbf{r}_{n+1}\sigma'}+h.c.\right)\right] \\
  S&=&\frac{\hbar^2}{m^*a_0^2}\sigma_0+E_Z(\cos{\phi}\sigma_x+\sin{\phi}\sigma_y)  \\
  T_n&=&-\frac{\hbar^2}{2m^*a_0^2}\sigma_0 + \frac{i}{2a_0} \frac{\mathbf{r}_{n+1}-\mathbf{r}_{n}}{|\mathbf{r}_{n+1}-\mathbf{r}_{n}|} \cdot (\beta\sigma_x+\alpha\sigma_y,-\beta\sigma_y-\alpha\sigma_x)
\end{eqnarray}
\end{widetext}
where the ${\sigma}$ and ${\sigma}'$ summations run on $\{\uparrow, \downarrow\}$, the up and down spin projections along $z$-axis.
The spin independent nearest neighbor hopping in the model is $t_h=\hbar^2/(2m^*a_0^2)$.

By choosing $a_0\approx \lambda_F/10$ we ensure parabolic behavior at
the working energy. The left and right leads are connected to site
$\mathbf{r}_{L}$ ($=\mathbf{r}_{N/2}$) and site $\mathbf{r}_{R}$
($=\mathbf{r}_{1}$), respectively.  We choose both leads well coupled
to the ring. This is achieved with the leads being simple
tight-binding chains with large bandwidth hopping $t_c\gg t_h$. The
Green function at the end site of each semi-infinite chain lead is
$g^r_c(E)\approx-i\sigma_0/|t_c|$, its real part being negligible
since we work at the center of the band assuring featureless energy
dependence. The self-energy of leads
$K=\{L,R\}$ are $\hat{\Sigma}^r_{K} (E) \approx -i
\left|\mathbf{r}_K\right>\left<\mathbf{r}_K\right|\sigma_0
t_i^2/|t_c|$ with the intermediate hopping $t_i$ fixed at the optimum
$t_i=(t_h+t_c)/2$. The conductance is computed using the retarded and
advanced Green functions $\hat{G}^r(E)=(\hat{G}^a(E) )^\dagger=[ (E +i
  0^+)I_N -\hat{H}-\hat{\Sigma}^r_{L}-\hat{\Sigma}^r_{R}]^{-1}$ with
\begin{equation}
G=\frac{e^2}{h} \text{Tr}\left[ \hat{\Gamma}_{R} (E_F)\hat{G}^r(E_F) \hat{\Gamma}_{L} (E_F)\hat{G}^a(E_F)\right],
\end{equation}
where the level-width function of lead $K$ is given by $\hat{\Gamma}_{K}(E)=i(\hat{\Sigma}^r_{K} (E)-\hat{\Sigma}^a_{K} (E))$.

The 1D tight-binding method does not take into account the large spin-dephased contribution that appear in 2D multi-mode
calculations with disorder. When comparing with other methods we add therefore a large isotropic background conductance $G_{\rm B}$
to the conductance that is independent of $\varphi$ as $G_{\Omega}(\varphi)=G_{\rm 1D,\Omega}(\varphi)+G_{\rm B}$.

\subsection{1D spin-rotation method}

\label{appendixspinrot}
We use a 1D single-mode approach to study spin-interference by introducing the effective spin Hamiltonian 
\begin{equation}
H_{\rm S}(\theta)=({\bf B}_{\rm R}+{\bf B}_{\rm D}+g^* \mu_{\rm B}{\bf B}_{\rm in-plane})\cdot {\bm \sigma},
\label{eq:eff_field}
\end{equation}
where the effective Rashba and Dresselhaus SO fields acting on a spin moving at constant speed are given by
\begin{equation}
{\bf B}_{\rm R}  = B_{\rm R} (\sin \theta\, \hat{\bf x} - \cos \theta\,  \hat{\bf y}),
\end{equation}
\begin{equation}
{\bf B}_{\rm D}  = {B}_{\rm D} ( -\cos \theta\,  \hat{\bf x} + \sin\theta\,  \hat{\bf y} ),
\end{equation}
respectively. Here $B_{\rm R}=k_{\parallel}\alpha$ and ${B}_{\rm D} =k_{\parallel}\beta$,
$k_{\parallel}$ is the wave number along the direction of propagation, and
$\theta$ is the direction of the $\bf k$-vector with ${\bf k}=k_{\parallel}(\cos \theta\,  \hat{\bf x} + \sin\theta\,  \hat{\bf y} )$.
Disorder in realistic rings suppresses multiple windings of spin around the ring and therefore 
we assume that AAS paths are the most important ones (Fig.~\ref{fig1}). As a consequence, resistance is determined  by backscattering of spins.

Using Eq.~(\ref{backscattering}), the backscattering probability amplitude is calculated by adding the contributions of the reflected wave at the intersection and the waves transmitted around the ring. The conductance is then given by
\begin{equation}
G=(2-p_r)G_0=(2-|\psi|^2)G_0,
\label{eq-conductance}
\end{equation}
where is $p_r=|\psi|^2$ is the backscattering probability. In pure 1D
models, reflected waves at the intersections may acquire a phase shift
of $\pi$.  This is associated with a negative value of $p_1$ in
Eq.~(\ref{backscattering}).  In 2D models, mode-mixing at the
intersections affects also phases of the waves transmitted around the
ring. In addition, disorder is assumed to give rise to spin dephasing
and relaxation. Therefore we keep $p_{1,2}$ as free parameters in the
1D spin-rotation model. At the intersection, a spin can move either
back to the lead or go around the ring in the clockwise/anticlockwise
direction. From this we can derive that $p_2 < \sqrt{(1-p_1^2)/2}$ in
general.

The spin evolution in a clockwise rotating round-trip around the ring is calculated from the unitary spin propagator
\begin{equation}
C=\prod_{m=1}^n\exp(iH_{\rm S}(\theta_m)\Delta t),
\label{propagator}
\end{equation}
where a full rotation ($\theta=\pi/2 \to -3\pi/2$ for the clockwise path in Fig.~\ref{fig1}) is divided into $n$ steps of
time interval $\Delta t$ and $\theta_m=\pi/2-2\pi m/n$.
The anticlockwise rotating round-trip operator $A$ is obtained likewise
by calculating the rotation in opposite direction ($\theta=-\pi/2 \to 3\pi/2$ for the anticlockwise path in Fig.~\ref{fig1}).
Details of the method are described in Ref.~\onlinecite{saarikoski2}.
From Eqs.~(\ref{backscattering}) and (\ref{eq-conductance}) we get a final expression for 
conductance of one spin-compensated incoming mode
\begin{eqnarray}
G=\big [&2-2p_1^{\;2}+2p_1^{\;2}p_2^{\;2}{\rm Tr} \big(C+C^{\dagger}+A+A^{\dagger}\big) \nonumber\\
&+p_2^{\;2}\big(4+{\rm Tr} (A^{\dagger}C+C^{\dagger}A)\big)]G_0.
\label{eq-conductance-aas}
\end{eqnarray}

\section{Resistance anisotropy in isotropic spin-orbit fields (case $\beta=0$)}

\label{isotropiccase}
As discussed in the main text, resistance anisotropy in our
calculations arises mostly due to spin-orbit field anisotropy. In the
absence of the Dresselhaus SO interaction, the rotation symmetry of
the Rashba SO field implies that the total average field acting on a
spin in a round-trip around the ring does not depend on the in-plane
field direction, so that resistance anisotropy is not expected. Still,
our calculations exhibit a small but nonvanishing resistance
anisotropy in isotropic spin-orbit fields as a function of the
relative orientation of the leads and the external in-plane magnetic
field, as we show below.

\begin{figure}
\includegraphics[width=\columnwidth]{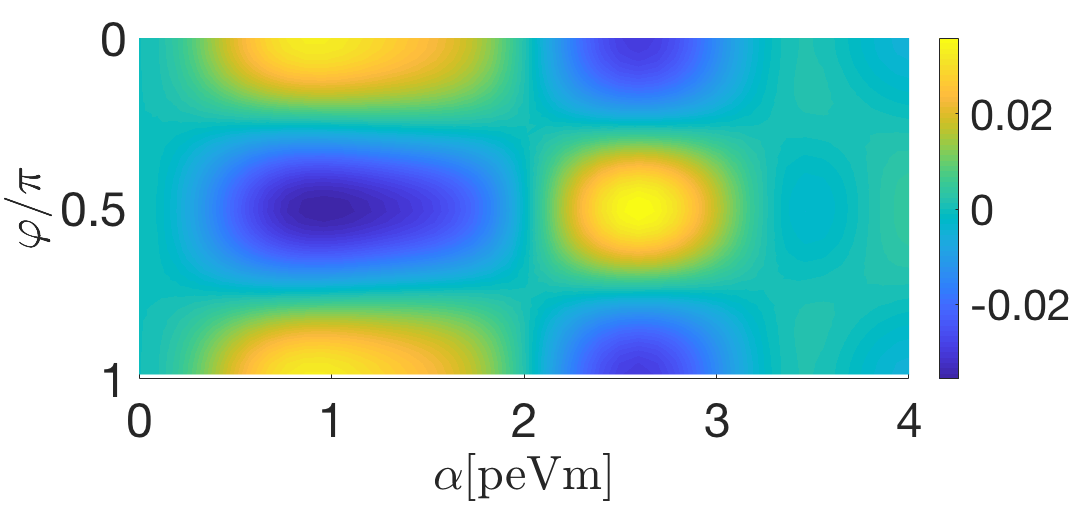}
\caption{Resistance anisotropy  Eq.~(\ref{averagedanisotropy}) (in \%)  in isotropic Rashba spin-orbit field
and vanishing Dresselhaus field $\beta=0\,{\rm peV m}$. The figure is calculated for the backscattered AAS paths ($p_1=0$, $p_2=0.035$)
with the 1D spin-rotation model as a function of in-plane field direction $\varphi$ and Rashba spin-orbit field $\alpha$.
Maximum anisotropy occurs between in-plane field axis $\varphi=0$ and $\varphi=\pi/2$.
\label{fig-isotropyaxis}}
\end{figure}

Figure~\ref{fig-isotropyaxis} shows resistance anisotropy calculated from the AAS backscattering probability in
a Rashba field using the 1D spin-rotation model.
The results show that at $\beta=0$ the resistance anisotropy is largest between in-plane fields parallel
($\varphi=0$) and normal ($\varphi=\pi/2$) 
to the lead direction, and vanishes between directions $\varphi=\pi/4$ and $3\pi/4$.

\begin{figure}
\includegraphics[width=\columnwidth]{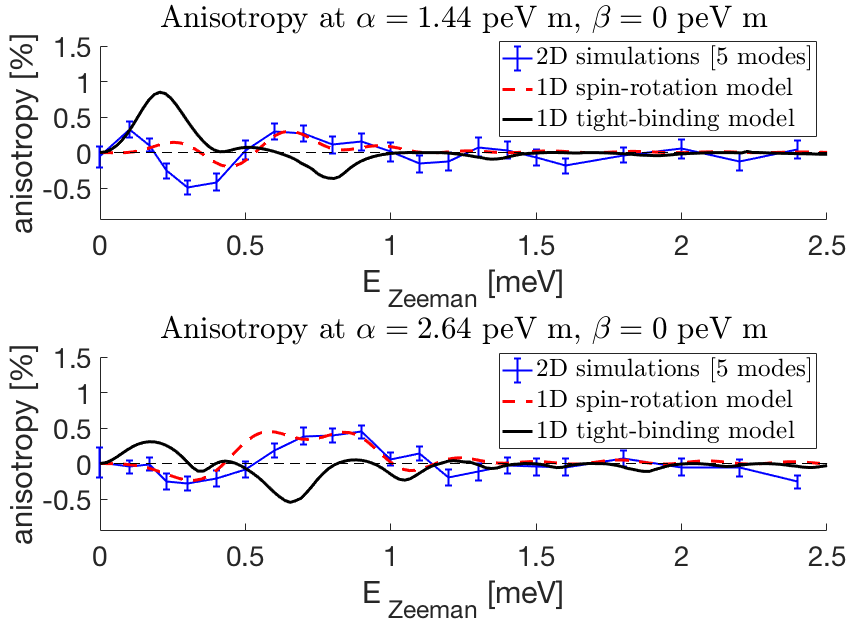}
\caption{Weak oscillations in resistance anisotropy of a $R=610\;{\rm nm}$ ring system as a function of in-plane field strength for isotropic Rashba spin-orbit fields $\alpha=1.44\,{\rm peV m}$ (above) and $\alpha=2.64\,{\rm peV m}$ (below) in the absence of Dresselhaus coupling ($\beta=0$). Resistance anisotropies are calculated between $\varphi=0$ and $\varphi=\pi/2$ axis
using Eq.~(\ref{angleanisotropy}).
Results are calculated using the 2D method in a $W=68$ nm wide wire with 5 transport modes and compared to the 1D spin-rotation
model. The 1D model assumes weak backscattering via AAS paths only ($p_1=0$, $p_2=0.035$).
\label{fig-beta0anisotropy}}
\end{figure}

Due to isotropic character of the Rashba SO field, the resistance anisotropy at $\beta=0$ is presumably more sensitive to effects of disorder.
Indeed, 2D calculations in 5-mode disordered systems show only a weak resistance anisotropy as
a function of the in-plane field direction (see Fig.~\ref{fig-beta0anisotropy}).
These oscillations in resistance anisotropy are in close agreement with the results obtained by using the 1D spin-rotation model
for AAS paths ($p_1=0$, $p_2=0.035$ shown in Fig.~\ref{fig-beta0anisotropy}).


\section{Zeeman oscillations in resistance}

\label{appendixzeeman}
\begin{figure}
\includegraphics[width=\columnwidth]{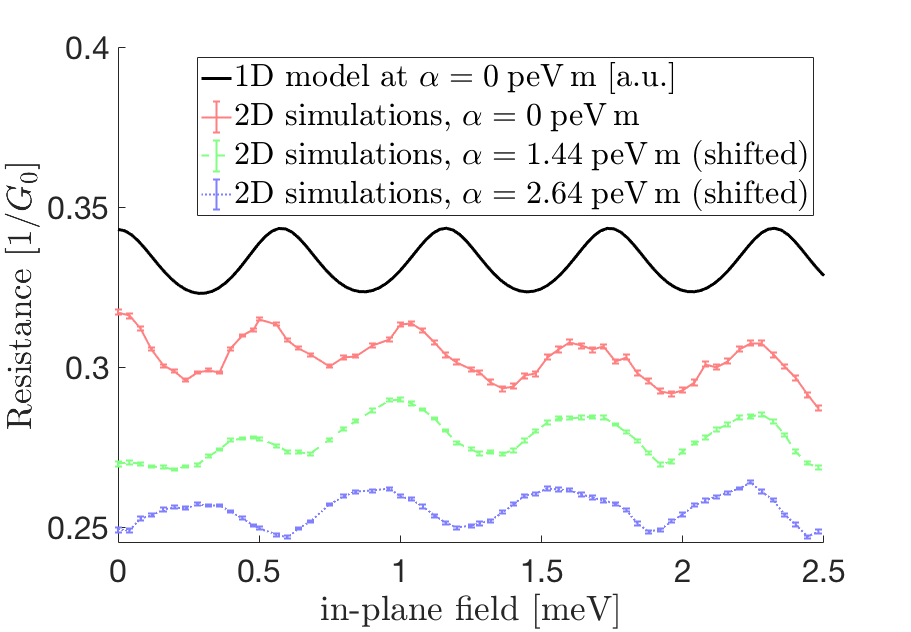}
\caption{Zeeman oscillations in resistance of a 5-mode mesoscopic
ring system calculated with the 2D method at Rashba
SO coupling strengths $\alpha=0\;{\rm peV\, m}$, $\alpha=1.44\;{\rm peV\, m}$, and
$\alpha=2.64\;{\rm peV\, m}$. The results are compared to the single-mode 1D spin-rotation model
at  $\alpha=0\;{\rm peV\, m}$ using $p_1=0.06$ and $p_2=0.1$ (thick line).
The Dresselhaus SO coupling is fixed at $\beta=0.3\;{\rm peV\, m}$ in all calculations. 
The mean free path of electrons in the 2D simulations is $\ell=2.5\;\mu {\rm m}$, the ring radius
is $0.61\;\mu{\rm m}$, and the in-plane magnetic field angle $\varphi=3\pi/4$.
The oscillations indicate interference between waves reflected at the intersection and waves transmitted around the ring.}
\label{zeemandiagram}
\end{figure}

The anisotropy oscillations as a function of the in-plane field are
related to Zeeman oscillations in the resistance associated with the
dynamic phase that the spins accumulate in a round-trip around the
ring.  These oscillations are especially prominent in loop
geometries.\cite{saarikoski} Since the dynamic phases acquired for
clockwise and counterclockwise rotating paths in symmetric ring geometries
are equal, Zeeman oscillations do not appear for directly transmitted (Aharonov-Casher) paths in
the absence of spin-orbit interaction.  However, Zeeman oscillations in resistance appear
due to interference between directly reflected and (counter)clockwise propagating paths.
Fig.~\ref{zeemandiagram} shows resistance calculated for AAS
paths with the 1D spin-rotation model and 2D simulations.  The
Dresselhaus term is fixed here at $\beta=0.3\,{\rm peV\,m}$.  The
Zeeman oscillation period is about 0.6 meV in the 1D model
corresponding to the Zeeman phase of $2\pi$ in a round-trip around the
610 nm radius ring.


\section{Supplementary data}

\label{supplydata}
We present here supplementary data associated with the topological transition in resistance anisotropy.

\begin{figure*}
\includegraphics[width=2\columnwidth]{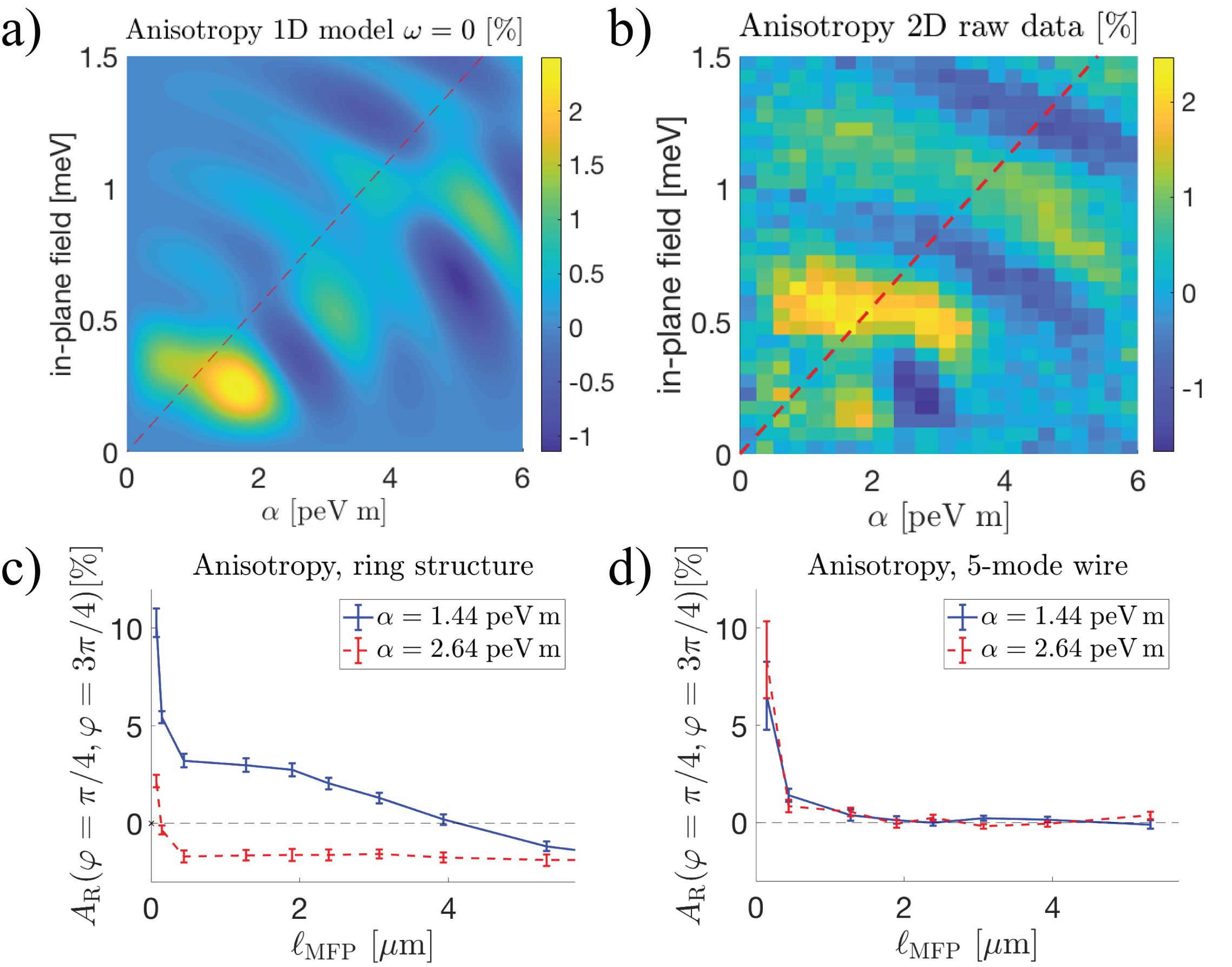}
\caption{a) Resistance anisotropy calculated from the backscattering probability for AAS paths using the 1D spin-rotation model
at fixed lead orientation along $x$-direction ($\omega=0$). 
b) Raw data of resistance anisotropy calculated using the 2D model with mean-free path of electrons $\ell=2.5\;{\mu}$m and $\omega=0$.
All anisotropies are calculated between in-plane field directions $\varphi_1=\pi/4$ and $\varphi_2=3\pi/4$ and
with Dresselhaus SO strength $\beta=0.3\;{\rm peV\,m}$ using Eq.~(\ref{angleanisotropy}).
The dashed line in the figures shows the critical line where the Rashba SO field strength is equal to the in-plane field strength.
c) Resistance anisotropy 
between  $\varphi=\pi/4$ and $\varphi=3\pi/4$ axis calculated
as a function of electron mean free path $\ell$. Here $\beta=0.3$ peV m and in-plane field strength of 0.17 meV.
The results are obtained using the 2D method with 5 transport modes in the ring.
d) { Resistance anisotropy between  $\varphi=\pi/4$ and $\varphi=3\pi/4$ axis in a straight 5-mode
wire (without the ring structure). The Dresselhaus SO strength and the Zeeman strength are the same as in c).
Anisotropy vanishes in the regimes of moderate and low disorder.}
\label{supply}}
\end{figure*}

Figure~\ref{supply}a shows the topological transition at a fixed lead direction $\omega=0$.
We note that lead direction averaging (Fig.~\ref{figy}b) evens out nonadiabatic effects close to the
critical line leaving a slightly more clear signature of the topological transition.
Figure~\ref{supply}b presents raw data of the 2D model associated with Fig.~\ref{figy}d.
Noise in the data is due to lattice disorder model.


{
Finally, we note that the resistance anisotropy persists when the
disorder density is increased from the value used in the main text
(corresponding to $\ell=2.5\;\mu {\rm m}$). Figure~\ref{supply}c shows
the anisotropy $A_{\rm R}(\varphi=\pi/4,\varphi=3\pi/4)$ for
$\beta=0.3$ peV m and in-plane field strength of 0.17 meV, {\em i.e.}
the same as values used in Fig.~\ref{fig2} and
Fig.~\ref{fig-anisotropyaxis-1D}, and 5 transport modes.  
Resistance anisotropy increases slightly with decreasing electron mean free path.
In the clean regime, $\ell \gg 3\;\mu
{\rm m}$, direct paths through the system dominate over AAS paths and
resistance anisotropy decreases. 

Figure~\ref{supply}c shows that the sign of $A_{\rm R}(\varphi=\pi/4,\varphi=3\pi/4)$
reverses in the higher Rashba SO coupling strength $\alpha=2.64\;{\rm peV\, m}$
at moderate disorder densities with $0.45\;{\rm \mu  m}<\ell<3.0\;{\rm \mu m}$.
This is consistent with AAS path interference effects described in the main text.
However, we observe no sign reversal in the highly disordered regime $\ell<0.15\;{\rm \mu  m}$.
Our calculations show that in this highly disordered regime,
resistance anisotropy originates mainly from interference effects within the multi-mode wire in
contrast to AAS ring interference. This is demonstrated by transport simulations in a straight 5-mode wire
without a ring structure in Fig.~\ref{supply}d.
In the 5-mode wire resistance anisotropy is not significant in the regime of moderate disorder
but increases with disorder strength when $\ell < 0.45\;{\rm \mu  m}$.
Moreover, anisotropy does not reverse sign at higher Rashba SO coupling $\alpha=2.64\;{\rm peV\, m}$.
This pattern can be directly compared to calculations in a ring structure in Fig.~\ref{supply}c.
We conclude that AAS ring interference is the most important source of resistance
anisotropy in the ring device in the regime of moderate disorder.}
\end{appendix}


\begin{thebibliography}{99}
\bibitem{footnote} In the absence of Dresselhaus SO interaction, the total average field acting on a spin in a round-trip is isotropic with respect to the in-plain field direction. Still, the resistance across the system is weakly anisotropic with respect to the orientation of the lead relative to that of the in-plane field. We discuss this effect in Appendix \ref{isotropiccase}.
\bibitem{footnote1} Smoothing has been performed on the data to reduce noise associated with the lattice disorder model.
The raw data is shown in Appendix~\ref{supplydata}.
\end{thebibliography}
\end{document}